\documentclass[10pt]{IEEEtran}
\usepackage{amsmath}
\usepackage[T1]{fontenc}
\usepackage{algorithm}
\usepackage{alphalph}
\usepackage{algorithmic}
\usepackage{listings}

\usepackage{nomencl}
\makenomenclature

\usepackage{amsmath}
\usepackage{graphicx}
\usepackage{color}
\usepackage[cmintegrals]{newtxmath}
\usepackage[caption=false,font=footnotesize]{subfig}
\ifCLASSOPTIONcompsoc
\usepackage[caption=false,font=normalsize,labelfont=sf,textfont=sf]{subfig}
\else
\usepackage[caption=false,font=footnotesize]{subfig}
\fi
\usepackage[colorlinks,citecolor=red,urlcolor=blue,bookmarks=false,hypertexnames=true]{hyperref} 
\usepackage{caption}
\usepackage{lipsum}
\usepackage{bbm}
\usepackage{nameref}
\usepackage{varioref}
\usepackage{hyperref}
\usepackage{setspace}

\newtheorem{lemma}{Lemma}
\newtheorem{proposition}{Proposition}

\begin{document}
\title{Cyber-Physical-Systems and Secrecy Outage Probability: Revisited}
\author{Makan~Zamanipour,~\IEEEmembership{Member,~IEEE}
\thanks{Manuscript received NOV, 2021; revised X XX, XXXX. Copyright (c) 2015 IEEE. Personal use of this material is permitted. However, permission to use this material for any other purposes must be obtained from the IEEE by sending a request to pubs-permissions@ieee.org. Makan Zamanipour is with Lahijan University, Shaghayegh Street, Po. Box 1616, Lahijan, 44131, Iran, makan.zamanipour.2015@ieee.org. }
}

\maketitle
\markboth{IEEE, VOL. XX, NO. XX, X 2021}%
{Shell \MakeLowercase{\textit{et al.}}: Bare Demo of IEEEtran.cls for Computer Society Journals}
\begin{abstract}
This paper technically explores the secrecy rate $\Lambda$ and a maximisation problem over the concave version of the secrecy outage probability (SOP) as $\mathop{{\rm \mathbb{M}ax}}\limits_{\Delta  } {\rm \; }   \mathbb{P}\mathscr{r} \big(     \Lambda \ge \lambda \big)  $. We do this from a generic viewpoint even though we use a traditional Wyner's wiretap channel for our system model $-$ something that can be extended to every kind of secrecy modeling and analysis. We consider a Riemannian mani-fold for it and we mathematically define a volume for it as $\mathbb{V}\mathscr{ol}\big \lbrace \Lambda \big \rbrace$. Through achieving a new bound for the Riemannian mani-fold and its volume, we subsequently relate it to the number of eigen-values existing in the relative probabilistic closure. We prove in-between some novel lemmas with the aid of some useful inequalities such as the \textit{Finsler’s} lemma, the generalised \textit{Young’s} inequality, the generalised \textit{Brunn-Minkowski} inequality, the \textit{Talagrand’s} concentration inequality. We additionally propose a novel Markov decision process based reinforcement learning algorithm in order to find the optimal policy in relation to the eigenvalue distributions $-$ something that is extended to a possibilisitically semi-Markov decision process for the case of periodic attacks.
\end{abstract}

\maketitle

\IEEEdisplaynontitleabstractindextext
\IEEEpeerreviewmaketitle

\section{Introduction} 

\IEEEPARstart{P}hysical-layer security inevitably plays a vital role in 5/6 G and beyond. This widely supported concept \cite{qaq1, qaq2, qaq3, q1, q2, q3, q4, q5, q6, q7, q8, q9} is emerged in parallel with traditional cryptography techniques while information-theoretic perspectives are promising. 

In order to simultaneously enhance the fairness and the quality of service among all the users, the physical characteristics of the wireless channel are of an absolutely inconsistent nature, which originally comes from the channel’s broadcast behaviour $-$ something that should be essentially managed. 

The concept of secrecy outage probability (SOP) in telecommunication still shows up an open research field in the literature. This concept is useful e.g. for: free-space optical communications \cite{qaq1}, vehicular communications \cite{qaq2}, reflecting intelligent surfaces \cite{qaq3}-\cite{q1}, cognitive networks \cite{q2}, cooperative communications \cite{q3}, power-line communications \cite{q4}, the internet of Things \cite{q5}, terrestrial networks \cite{q6}, mobile edge computing networks \cite{q7}, molecular communications \cite{q8} and under-water networks \cite{q9}.

In \cite{qaq1}-\cite{q9} and in totally various types of system models, some novel and closed-form mathematical expressions have been newly derived and proposed $-$ some of them are optimisation based, some of them are statistical oriented and some of them are even jointly theoretical-practical.

\subsection{Motivations and contributions} In this paper, we are interested in responding to the following question: \textit{How can we guarantee highly adequate relaxations over the principle of SOP?} With regard to the non-complete version of the literature, the expressed question strongly motivate us to find an interesting solution, according to which our contributions are fundamentally described as follows. 
\begin{itemize}
\item \textcolor{red}{\textbf{(\textit{i})}} A new bound in relation to the maximisation problem over the SOP's concave version is derived. We, in addition, theoretically discuss about a totally novel interpretation over the aforementioned maximisation problem from a duality point of view. We consider a Riemannian mani-fold for the SOP's concave version and we mathematically define a volume for it for which we derive a new bound. We use some insightful principles such as Keyhole contour.

\item \textcolor{red}{\textbf{(\textit{ii})}} We subsequently relate the Riemannian mani-fold and its bounded volume expressed above to the number of eigen-values. We use in-between some useful lemmas and inequalities such as the Finsler’s lemma, the generalised Young’s inequality, the generalised Brunn-Minkowski inequality, the Talagrand’s concentration inequality. 

\item \textcolor{red}{\textbf{(\textit{iii})}} We additionally go over further discussions in terms of the \textit{projection method} technically relating it to the former parts. We also take into account the case of relaxing the non-contractibility and how to decrease the relative non-contractibility radius from a topological point of view. In this context, we propose a novel Markov decision process based reinforcement learning algorithm in order to find the optimal policy in relation to the eigenvalue distributions $-$ something that was subsequently extended to a semi-Markov decision one for the case of periodic attacks and with regard to the possibility-theory.

\end{itemize}

\begin{figure}[t]
\centering
\subfloat{\includegraphics[trim={{47mm} {44 mm} {23mm} {53mm}},clip,scale=0.4]{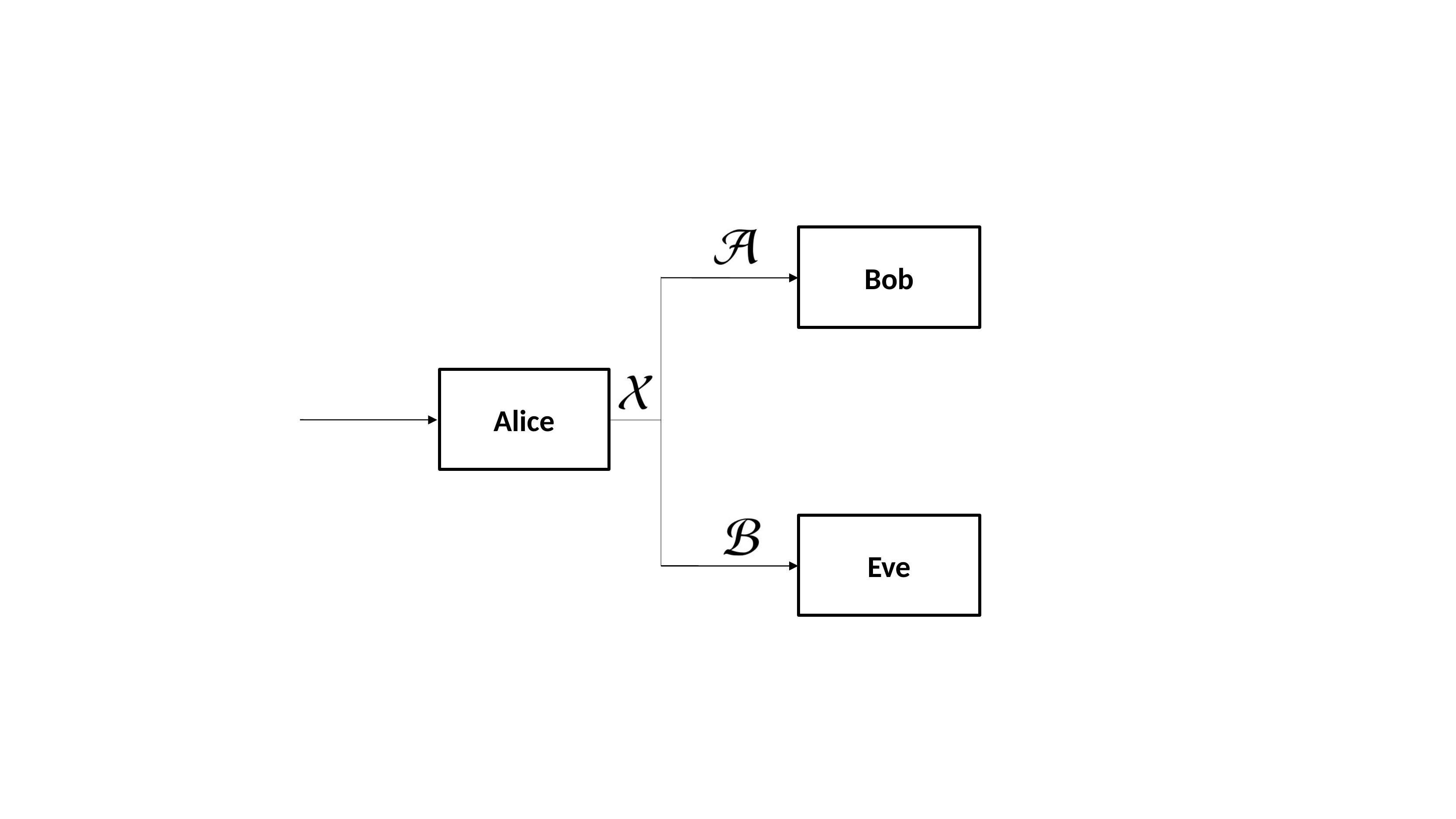}}
\caption{A traditional Wyner's wiretap channel: System diagram of a communication which suffers from insecurity \textcolor{black}{while $\mathcal{X}$, $\mathcal{A}$ and $\mathcal{B}$ are random states relating to respectively Alice, Bob and Eve.} }
\label{F3}
\end{figure} 

\subsection{General notation} The notations widely used throughout the paper is given in Table \ref{table1}.

\begin{table*}[h]
\begin{center}
\captionof{table}{{List of notations.}}
 \label{table1} 
\begin{tabular}{p{2.091cm}|p{3.45cm} |p{3.26cm}|p{3.84cm}}
 \hline      
 \hline
Notation        &  Definition  & Notation          &  Definition   \\
\hline
    $\mathbb{M}in$   &                   Minimisation    &           $\mathbb{M}ax$    &     Maximisation          \\
           $\mathbb{E}$ & Expected-value   &      $\mathcal{I} (\cdot)$      &  Mutual-information      \\
$\stackrel{def}{=} $         &     Is defined as      &      $\approx$      &   Is approximated to     \\
           $ \mathbb{V}\mathscr{ol}$    &    Volume       &      $\mathbb{P}\mathscr{r}  $    &       Probability               \\
     $det(\cdot)$      &    Matrix determinant       &      $(\cdot)^{T}$      &    Transpose    \\
            $Tr [\cdot]$    &     Trace of matrix      &      $(\cdot)^{-1}$      &    Inverse of matrix    \\
$\Lambda $ & Secrecy rate & $ \mathbb{P}\mathscr{r} \big(     \Lambda \ge \lambda \big)$ & SOP's Concave Version\\
$\mathscr{t}$ & Time & $\mathscr{H}$  & Entropy\\
$\mathcal{M}$ &Riemannian-manifold&$\partial \mathcal{M}$& Boundary\\
$\zeta$& Eigen-value & $Tr[\cdot]$ &Trace \\
\hline
 \hline
\end{tabular}
\end{center}
\end{table*}

\subsection{Organisation}
\begin{figure*}[t]
\centering
\subfloat{\includegraphics[trim={{17mm} {21.7 mm} {1mm} {11mm}},clip,scale=0.6]{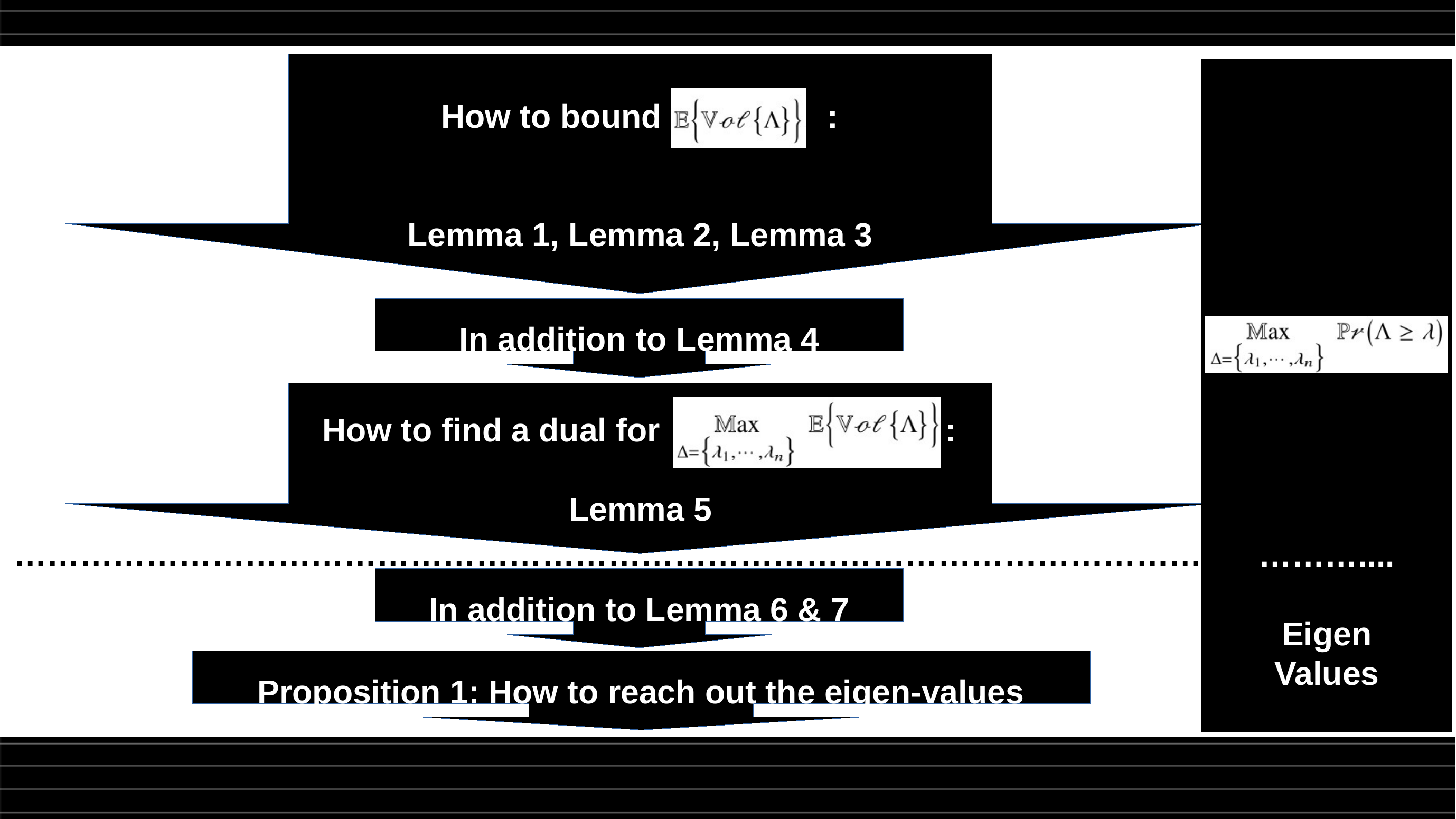}}
\caption{\textcolor{black}{Flow of problem-and-solution.}}
\label{F4}
\end{figure*}

The rest of the paper is organised as follows. The system set-up and our main results are given in Sections II and III. Subsequently, the evaluation of the framework and conclusions are given in Sections IV and V. \textcolor{black}{In Fig. \ref{F4}, the flow of the main problem and the solution to that is deppicted. }

\section{System model and problem formulation}
In this section, we describe the system model, subsequently, we formulate the basis of our problem. 

\subsection{System description: A traditional Wyner's wiretap channel $-$ without loss of generality}

A traditional Wyner's wiretap channel\footnote{Although our novel analysis can be undoubtedly extended to other scenarios as well. For example, for a reconfigurable intelligent surface based scheme, the lower-bound of the secret-key-rate as the maximal key bits generated from
an observation is expressed as \cite{papa}: $\mathop{{\rm \mathbb{M}ax}} \Big \lbrace  \big \lbrace \mathcal{I} \big ( h_{ab}; h_{ba}  \big)-\mathcal{I} \big ( h_{ab}; h_{ae}, h_{be}  \big)   \big \rbrace ,   \big \lbrace \mathcal{I} \big ( h_{ba}; h_{ab}  \big)-\mathcal{I} \big ( h_{ba}; h_{ae}, h_{be}  \big) \big \rbrace\Big \rbrace$ while $(\cdot)_{ab}$, $(\cdot)_{ba}$, $(\cdot)_{ae}$, $(\cdot)_{be}$ respectively declare the links Alice-to-Bob, Bob-to-Alice, Alice-to-Eve and Bob-to-Eve and $h$ stands for the stacked versions of measurements at the relative receiver. Nevertheless, the physical logic behind the aforementioned rate here is similar to our current consideration and scheme, and we have nothing to to with their detail since we are supposed to find a relaxation over the outage probability relating to the our security oriented rates.} based communication scenario includes a transmitter named Alice and a legitimate reciever named Bob and an un-authorised one as an eavesdropper named Eve\textcolor{black}{, as shown in Fig. \ref{F3}}. The information capacity of the communication system is theoretically expressed by the general formula from \textit{Shannon}. The secrecy capacity is interpreted as a bound of the security performance of the communication system. We now have the following inequality \textcolor{black}{for the secrecy capacity from an information-theoretic point of view}
\begin{equation}
\begin{split}
\;C_{s} \stackrel{def}{=} \mathop{{\rm \mathbb{M}ax}}\limits_{f_{\mathcal{X}}( x) } {\rm \; } \mathcal{I} \big ( \mathcal{X},\mathcal{A}\big)-\mathcal{I} \big ( \mathcal{X},\mathcal{B}\big) \;\;\;\;\;\;\;\;\;  \\
\ge \mathop{{\rm \mathbb{M}ax}}\limits_{f_{\mathcal{X}}( x) } {\rm \; } \mathcal{I} \big ( \mathcal{X},\mathcal{A}\big)-\mathop{{\rm \mathbb{M}ax}}\limits_{f_{\mathcal{X}}( x) } {\rm \; } \mathcal{I} \big ( \mathcal{X},\mathcal{B}\big),
\end{split}
\end{equation}
while $\mathcal{X}$, $\mathcal{A}$ and $\mathcal{B}$ are random states relating to respectively Alice\footnote{\textcolor{black}{Encoded by Alice.}}, Bob\footnote{\textcolor{black}{Observed by Bob.}} and Eve\footnote{\textcolor{black}{Observed by Eve.}}, \textcolor{black}{where the maximisation takes place over the encoding function $f_{\mathcal{X}}( x) $ $-$ and consequently input distributions\footnote{\cite{tt}.} $-$ , that is, $f_{\mathcal{X}}( x) $ $-$ the input distributions $-$ should be optimally found by Alice in the sense that the overall performance can be technically guaranteed with regard to the metric of the secrecy capacity\footnote{\textcolor{black}{For the precise definition of rate and capacity, please refer e.g. to \cite{tt}.}} $C_{s}$. Hereinafter, we r-call $C_{s}$ as $\Lambda$.}

\subsection{Main problem}
\textcolor{black}{The main problem w.r.t. the secrecy rate $\Lambda$ is about the maximisation problem over the SOP's concave version $   \mathbb{P}\mathscr{r} \big(     \Lambda \ge \lambda \big)  $, that is, $\mathop{{\rm \mathbb{M}ax}}\limits_{\Delta = \big \lbrace \lambda_1, \cdots, \lambda_n\big \rbrace  } {\rm \; }   \mathbb{P}\mathscr{r} \big(     \Lambda \ge \lambda \big)  $ where the parameters are defined in the next section. Furthermore, we discuss how to reach out the relative eigen-values.}

\textsc{\textbf{Assumption 1.}} \textit{We consider that the dynamical system}
\begin{equation}
\begin{split}
\; \mathcal{X}^{(t+1)}=a_1\mathcal{X}^{(t)}+\mathscr{w}_0^{(t+1)}, \;\;\;\;\;\;\;\;\;\;\;\;\;\;\;\;\;\;\;\;\;\;\;\;\;\;\\
\big(  \mathcal{A}^{(t)},\mathcal{B}^{(t)}  \big)=\big(a_2, a_3,\big)\mathcal{X}^{(t)}+\big(\mathscr{w}_1^{(t)}, \mathscr{w}_2^{(t)}\big),
\end{split}
\end{equation}
\textit{is satisfied while $(a_1,a_2, a_3)$ and $(\mathscr{w}_0,\mathscr{w}_1, \mathscr{w}_3)$ are respectively the control parameter tuple and the noise one.}
\textit{is unstable, that is, its spectral radius is}
\begin{equation}
\begin{split}
\; \varphi (a_1)=\mathop{{\rm \mathbb{M}ax}}\limits_{i } {\rm \; }|\zeta(a_1)|>1, 
\end{split}
\end{equation}
\textit{is satisfied.}

\section{Main results}

In this section, our main results are theoretically provided in details.

\textsc{\textbf{Definition 1.}} \textit{Let us theoretically assign a random variable $\Lambda(\mathscr{t})\stackrel{def}{=}\big \lbrace \lambda_1, \cdots, \lambda_n\big \rbrace$ for the secrecy rate according to which the SOP's concave version can be defined as $   \mathbb{P}\mathscr{r} \big(     \Lambda \ge \lambda \big)  $ as well.}

\begin{lemma} \label{P2} \textit{For the random variable $\Lambda(\mathscr{t})\stackrel{def}{=}\big \lbrace \lambda_1(\mathscr{t}), \cdots, \lambda_n(\mathscr{t})\big \rbrace$ over the time horizon $-$ for which the term $(\mathscr{t})$ is neglected hereinafter for the ease of notation $-$, the expression }
\begin{equation}
\begin{split}
\; \mathbb{E} \big \lbrace \Lambda  \big \rbrace \le \frac{\mathbb{E} \big \lbrace  e^ { \mathscr{t} \Lambda  } \big \rbrace-1}{\mathscr{t}}, 
\end{split}
\end{equation}
\textit{is satisfied.}\end{lemma} 

\textbf{\textsc{Proof:}} The proof is convenient to follow according to the Taylor-expansion theorem, however, see Appendix \ref{sec:A} for more detailed justifications.$\; \; \; \blacksquare$

In the following, roughly speaking, we consider the secrecy rate as a Riemannian mani-fold for which a volume is technically defined.

\textsc{\textbf{Remark 1:}} \textit{\textcolor{black}{The reason why we consider a vloume can be theoretically justified as follows. Unhesitatingly, since we consider the relative parameters as Riemannian mani-folds, one can justify a volume over them from a generic point of view. In particular and as obvious e.g. from \cite{ddd1, ddd2} and in the context of the \textit{Asymptotic Equipartition Property}, one can interpret the volume of a random variable $-$ only for the secrecy rate $\Lambda$ but not the SOP's concave version $   \mathbb{P}\mathscr{r} \big(     \Lambda \ge \lambda \big)  $ $-$ as the exponential of Shannon’s entropy of it\footnote{That is, $  \mathbb{V}\mathscr{ol}\big \lbrace  \Lambda (\mathscr{t})  \big \rbrace \rightarrow e^{     \mathscr{H} \big \lbrace  \Lambda  \big \rbrace}$ holds. In other words, $  \mathbb{V}\mathscr{ol}\big \lbrace  \Lambda (\mathscr{t})  \big \rbrace \approx  \mathbb{E}  \Big( \mathcal{I} \big(    \Lambda \big)\Big)$.}. Indeed, this property can be elaborated via the \textit{Cramer's large deviation theorem}\footnote{\textcolor{black}{It states that the probability of a large deviation from mean decays exponentially with the number of samples}. See the \textit{large deviations theory}.}. Meanwhile, in relation to the volume of the random variable, please do not misunderstand it with the standard deviation as we are generalising it. }}

\textsc{\textbf{Definition 2.}} \textit{Let us, \textcolor{black}{for more generalisations}, consider the secrecy rate as the Riemannian mani-fold $\mathcal{M}$ for which the volume $\mathbb{V}\mathscr{ol}\big \lbrace  \Lambda  \big \rbrace  $ is valid. Undoubtedly, $\mathcal{M}$ can be contractible if its Euler characteristics gets $\mathscr{X}(\mathcal{M})=1$\footnote{See e.g. \cite{GIL, KHAYEH20, LEE1} to understand what it is.}, that is, if $\mathcal{M}$ is homotopically\footnote{In \textit{topology}, two continuous functions from one topological space to another one $-$ whether isolated or not $-$ are called \textit{homotopic}.} equivalent to a single-point, that is, when and if $\mathcal{M}$ is continously shrunk and topologically deformed into a sigle-point. In other words, a space is contractible iff the identity map from it to itself $-$ which is always a homotopy equivalence $-$ is null-homotopic \cite{GIL, KHAYEH20, LEE1, LEE2}.}

In order to further justify Definition 2, re-call Assumption 1. Indeed, when a system is unstable, the  more uncertainty is amplified originating from the noise. Thus, Eve cannot always predict the system close to an equilibrium \cite{papas}. Or in other words, if: (\textit{i}) $(a_1,a_2)$ is observable; and (\textit{ii}) $(a_1,\sqrt{a_2})$ is controllable, the error covariance matrix at Eve converges exponentionally to a unique fixed-point \cite{Anderson}. We have consequently to consider inverse. 

\textbf{\textsc{Definition 3 $-$ Tangent Cone\footnote{See e.g. \cite{usd1, usd2, usd3} to understand what it is.}.}} \textit{The term tangent cone is defined as $ \omega_{\mathcal{M}}(x)\stackrel{def}{=} \lim_{h \to 0} inf \frac{dist \big(   x+h\omega,\mathcal{M} \big)}{h} =0 $ while $dist(\cdot,\cdot)$ stands literally for the information theoretic distance(s).}

\textbf{\textsc{Definition 4 $-$ Inward-pointing\footnote{See e.g. \cite{GIL, KHAYEH20, LEE1} to understand what it is.}.}} \textit{If $f(\cdot)$ is a vector-field relating to $ \mathcal{M}$ and the boundary $\partial \mathcal{M}$ exists, $f(\cdot)$ is said to be oint inward to $ \mathcal{M}$ at a point $x \in  \mathcal{M}$ if $f(x) \in  \omega_{\mathcal{M}}(x) - \big \lbrace  \omega_{\partial\mathcal{M}}(x)  \big  \rbrace$ holds\footnote{This is more-and-less similar to \textit{Pincare-Hopf-Theorem} in system theory and stabilizations.}.}

\textsc{\textbf{Corollary 1.}} \textit{Any compact and convex set is contractible, but not vise versa\footnote{See \cite{DOOD}, Page 4.}. Meamwhile, every contractible set, even non-convex, has a concave volume\footnote{See \cite{DOOD}, Page 4.}. Finally, every vector field that is inward-point to $\mathcal{M}$ has an equilibrium \cite{SHIRU1, SHIRU2} as there exists at least one contractible interior within the boundary $\partial \mathcal{M}$, if $\partial \mathcal{M}$ exists.}

\textsc{\textbf{Objective.}} \textit{Our aim is to increase the contractibility radius\footnote{See e.g. \cite{GIL} to understand what it is.} as much as possible.}

\textcolor{black}{\subsection{How to define a maximisation problem over the SOP's concave version  as $\mathop{{\rm \mathbb{M}ax}}\limits_{\Delta = \big \lbrace \lambda_1, \cdots, \lambda_n\big \rbrace  } {\rm \; }   \mathbb{P}\mathscr{r} \big(     \Lambda \ge \lambda \big)  $ w.r.t. the secrecy rate $\Lambda$}}

\begin{lemma} \label{P2} \textit{Vitale's random Brunn-Minkowski inequality\footnote{Generalised Brunn-Minkowski inequality \cite{1}.} $-$ The expression }
\begin{equation}
\begin{split}
\; \mathbb{V}\mathscr{ol}  \Big \lbrace           \mathbb{E} \big \lbrace  \Lambda  \big \rbrace   \Big \rbrace \ge \mathbb{E}  \Big \lbrace         \mathbb{V}\mathscr{ol}\big \lbrace  \Lambda  \big \rbrace   \Big \rbrace,
\end{split}
\end{equation}
\textit{holds.}\end{lemma} 

\begin{lemma} \label{P2} \textit{The expression }
\begin{equation}
\begin{split}
\; \mathbb{V}\mathscr{ol}  \Big \lbrace              \frac{\mathbb{E} \big \lbrace e^ { \mathscr{t} \Lambda  } \big \rbrace  -1}{\mathscr{t}}     \Big \rbrace \ge \mathbb{E}  \Big \lbrace         \mathbb{V}\mathscr{ol}\big \lbrace  \Lambda   \big \rbrace   \Big \rbrace,
\end{split}
\end{equation}
\textit{holds.}\end{lemma} 


\textbf{\textsc{Proof:}} The proof is easy to follow by an integration of \textit{Lemma 1} and \textit{Lemma 2}. \textcolor{black}{This is due to the fact that $\mathbb{E}  \Big \lbrace         \mathbb{V}\mathscr{ol}\big \lbrace  \Lambda   \big \rbrace   \Big \rbrace \le \mathbb{V}\mathscr{ol} \Big \lbrace       \underbrace{\mathbb{E}        \big \lbrace  \Lambda   \big \rbrace  }_{\le   \frac{\mathbb{E} \big \lbrace e^ { \mathscr{t} \Lambda  } \big \rbrace  -1}{\mathscr{t}}     }  \Big \rbrace$ holds}.$\; \; \; \blacksquare$

\begin{lemma} \label{P2} \textit{The expression }
\begin{equation}\label{eq:5}
\begin{split}
\; \mathbb{E} \big \lbrace  e^ { \mathscr{t} \Lambda  } \big \rbrace=e^ { \mathscr{t} \lambda} \mathbb{P}\mathscr{r} \big(     \Lambda \ge \lambda \big),
\end{split}
\end{equation}
\textit{strongly holds.}\end{lemma} 

\textbf{\textsc{Proof:}} See Appendix \ref{sec:B}.$\; \; \; \blacksquare$

\begin{lemma} \label{P2} \textit{The problem }
\begin{equation}
\begin{split}
\; \mathop{{\rm \mathbb{M}ax}}\limits_{\Delta = \big \lbrace \lambda_1, \cdots, \lambda_n\big \rbrace  } {\rm \; }  \mathbb{V}\mathscr{ol}  \Big \lbrace              e^ { \mathscr{t} \lambda}  \Big \rbrace \mathbb{V}\mathscr{ol}  \Big \lbrace\mathbb{P}\mathscr{r} \big(     \Lambda \ge \lambda \big)     \Big \rbrace,
\end{split}
\end{equation}
\textit{can be\footnote{Not definitely, but in terms of one of the highly probably efficient and acceptable one: See \cite{khar} for more details.} a dual one for the problem }
\begin{equation}
\begin{split}
\;  \mathop{{\rm \mathbb{M}ax}}\limits_{\Delta = \big \lbrace \lambda_1, \cdots, \lambda_n\big \rbrace  } {\rm \; } \mathbb{E}  \Big \lbrace         \mathbb{V}\mathscr{ol}\big \lbrace  \Lambda   \big \rbrace   \Big \rbrace.
\end{split}
\end{equation}
\end{lemma} 

\textbf{\textsc{Proof:}} See Appendix \ref{sec:C}.$\; \; \; \blacksquare$

\begin{algorithm*}
\caption{\textcolor{black}{A greedy algorithm to $  \Lambda$. }}\label{f}
\begin{algorithmic}
\STATE \textbf{\textsc{Initialisation.}} 

\textbf{while} $ \mathbb{TRUE}$ \textbf{do} 

$\;\;\;\;\;\;\;\;\;\;$Find the solution to $\Delta \in \Bigg \lbrace  \mathop{{\rm \mathbb{M}ax}}\limits_{\Delta = \big \lbrace \lambda_1, \cdots, \lambda_n\big \rbrace  } {\rm \; } \mathbb{E}  \Big \lbrace         \mathbb{V}\mathscr{ol}\big \lbrace  \Lambda   \big \rbrace   \Big \rbrace, \mathop{{\rm \mathbb{M}ax}}\limits_{\Delta = \big \lbrace \lambda_1, \cdots, \lambda_n\big \rbrace  } {\rm \; }   \mathbb{V}\mathscr{ol}  \Big \lbrace\mathbb{P}\mathscr{r} \big(     \Lambda \ge \lambda \big)     \Big \rbrace \Bigg \rbrace$. 

\textbf{endwhile} 

\textbf{end}

\textbf{\textsc{Output:}} $  \Lambda$
\end{algorithmic}
\end{algorithm*}

\textsc{\textbf{Remark 2:}} 
\begin{itemize}
\item (\textit{i}) $\mathbb{V}\mathscr{ol}  \Big \lbrace              e^ { \mathscr{t} \lambda}  \Big \rbrace.$ \textit{Whether $\mathop{{\rm \mathbb{M}ax}}\limits_{\Delta = \big \lbrace \lambda_1, \cdots, \lambda_n\big \rbrace  } {\rm \; }  \mathbb{V}\mathscr{ol}  \Big \lbrace              e^ { \mathscr{t} \lambda}  \Big \rbrace$ is of a partially useless nature here for our main problem or not, we use it as a trick which is of a purely useful nature in the next lemma.}
\item (\textit{ii}) $\mathbb{V}\mathscr{ol}  \Big \lbrace\mathbb{P}\mathscr{r} \big(     \Lambda \ge \lambda \big)     \Big \rbrace$. \textit{First of all, we have nothing to do with its maximum version, i.e., $1$. Additionally, recalling Definition 2 as well as Remark 1 in connection with $\mathbb{V}\mathscr{ol}\big\{ \cdot \big \}$, we see that the aforementioned value is not by-default equated with $1$.}
\item (\textit{iii}) \textit{\textsc{\textbf{Recaptulation.}}} So far, we have indeed recasted the problem $\mathop{{\rm \mathbb{M}ax}}\limits_{\Delta = \big \lbrace \lambda_1, \cdots, \lambda_n\big \rbrace  } {\rm \; }   \mathbb{P}\mathscr{r} \big(     \Lambda \ge \lambda \big)  $ into two parts, i.e., $\mathop{{\rm \mathbb{M}ax}}\limits_{\Delta = \big \lbrace \lambda_1, \cdots, \lambda_n\big \rbrace  } {\rm \; } \mathbb{V}\mathscr{ol}  \Big \lbrace              e^ { \mathscr{t} \lambda}  \Big \rbrace$ and $\mathop{{\rm \mathbb{M}ax}}\limits_{\Delta = \big \lbrace \lambda_1, \cdots, \lambda_n\big \rbrace  } {\rm \; } \mathbb{V}\mathscr{ol}  \Big \lbrace\mathbb{P}\mathscr{r} \big(     \Lambda \ge \lambda \big)     \Big \rbrace$ which are deterministic ones.
\end{itemize}

\begin{lemma} \label{P2} \textit{The problem }
\begin{equation}
\begin{split}
\; \mathop{{\rm \mathbb{M}ax}}\limits_{\Delta = \big \lbrace \lambda_1, \cdots, \lambda_n\big \rbrace  } {\rm \; }  \int \mathbb{V}\mathscr{ol}  \Big \lbrace              e^ { \mathscr{t} \lambda}  \Big \rbrace + \int \mathbb{V}\mathscr{ol}  \Big \lbrace\mathbb{P}\mathscr{r} \big(     \Lambda \ge \lambda \big)     \Big \rbrace,
\end{split}
\end{equation} 
\textit{can be a dual for the problem}
\begin{equation}
\begin{split}
\; \mathop{{\rm \mathbb{M}ax}}\limits_{\Delta = \big \lbrace \lambda_1, \cdots, \lambda_n\big \rbrace  } {\rm \; }  \mathbb{V}\mathscr{ol}  \Big \lbrace              e^ { \mathscr{t} \lambda}  \Big \rbrace \mathbb{V}\mathscr{ol}  \Big \lbrace\mathbb{P}\mathscr{r} \big(     \Lambda \ge \lambda \big)     \Big \rbrace,
\end{split}
\end{equation}
\textit{as its bound}.
\end{lemma}

\textsc{\textbf{Proof:}} The proof is easy to follow with the aid of the generalised Yong's ineqality\footnote{See e.g. \cite{e} to understand what it is.} which says that 
\begin{equation}
\begin{split}
\; f^{\prime}(x)g^{\prime}(x) \le  f(x)+ g(x),
\end{split}
\end{equation}
holds for the arbitary functions $f(\cdot)$ and $g(\cdot)$, while $(\cdot)^{\prime}$ stands for the drivative.$\; \; \; \blacksquare$

\textcolor{black}{\subsection{How to reach out the eigen-values}}

\textsc{\textbf{Remark 3:}} \textit{Regarding to the fact that mainly most of the secrecy rate problems can be discussed in the context of semi-definite algebra \cite{q1}-\cite{q9}, that is the format $\mathcal{B}^{T}\mathcal{A}\mathcal{B}$, we jump in terms of the following to the next steps.}

\begin{lemma} \label{P2} \textit{Finsler's lemma\footnote{See e.g. \cite{ee} to understand what it is.} $-$ The problem }
\begin{equation}
\begin{split}
\;  \exists \mathcal{X}, \mathcal{X}^{T}\mathcal{A}\mathcal{X}=\xi, \mathcal{X}^{T}\mathcal{B}\mathcal{X} \le \xi \Longrightarrow \exists z: \mathcal{B}-z \mathcal{A} < \xi , 
\end{split}
\end{equation}
\textit{holds for the arbitary matrices $\mathcal{A}$ and $\mathcal{B}$ while $\xi$ and $T$ stand respectively for an arbitary threshold and the transpose operand.}
\end{lemma}

\begin{proposition} \label{P2} \textit{Let us assume the descriptor system $\big( \mathcal{B}, \mathcal{A} \big)$, so, the characteristic polynomial is given as}
\begin{equation}
\begin{split}
\;  \mathscr{P}(z)=det\big( \mathcal{B}-z \mathcal{A} \big), 
\end{split}
\end{equation}
\textit{while $det(\cdot)$ stands for the matrix determinant. The number of eigenvalues in the region associated with the polynomial $\mathscr{P}(z)$ over the Riemannian $\mathbb{V}\mathscr{ol}  \Big \lbrace\mathbb{P}\mathscr{r} \big(     \Lambda \ge \lambda \big)     \Big \rbrace$ is related to $\big( \mathcal{B}-z \mathcal{A} \big)^{-1}$ and $det\big( \mathcal{B}-z \mathcal{A} \big)$ while $(\cdot)^{-1}$ stands for the inverse matrix.}
\end{proposition}

\textbf{\textsc{Proof:}} See Appendix \ref{sec:D}.$\; \; \; \blacksquare$

\subsection{Futher discussion}

\begin{proposition} \label{P2} \textit{Let the random $k-$dimensional subspace $\mathbb{P}\mathscr{r} \big \lbrace \Lambda \ge \lambda   \big \rbrace$ be valid and let $v^{\prime}_i$ be the projection of the point $v_i \in \Lambda$ into $\mathbb{P}\mathscr{r} \big \lbrace \Lambda  \ge \lambda  \big \rbrace$. Now, calling $L=||v^{\prime}_i-v^{\prime}_j||^2$ and $\tau_1=k||v_i-v_j||^2$. The value of $\mathbb{P}\mathscr{r} \big( L \leq (1-\tau_2)\tau_1  \big )$ is bounded from above by $\leq exp \big (  -\frac{k\tau^2_2}{4} \big), \forall \; 0< k < \infty$.}
\end{proposition}

\textbf{\textsc{Proof:}} See Appendix \ref{sec:E}.$\; \; \; \blacksquare$

\begin{proposition} \label{P2} \textit{Consider $\Theta \in \mathbb{R}^{r \times m}$ and $\theta \in \mathbb{R}^{r \times m}$. The element-wise projection operator $Proj(\cdot): \in \mathbb{R} \times \mathbb{R} \rightarrow \mathbb{R} $ which is a convex and continuously differentiable function is defined as \cite{cvcv}}
\begin{equation}\label{eq:13}
\begin{split}
\; Proj(\theta_{ij},\Theta_{ij})=\;\;\;\;\;\;\;\;\;\;\;\;\;\;\;\;\;\;\;\;\;\;\;\;\;\;\;\;\;\;\;\;\;\;\;\;\;\;\;\;\;\;\;\;\;\;\;\;\;\;\;\;\;\;\;\;\;\;\;\;\;\\\begin{cases}\Theta_{ij}-\Theta_{ij}f(\theta_{ij}), \;if \;f(\theta_{ij})>0\; \&\; \Theta_{ij} (\frac{df(\theta_{ij})}{d\theta_{ij}})>0,\\ \Theta_{ij}, o.w.,\end{cases}
\end{split}
\end{equation}
\textit{where the index $(\cdot)_{ij}$ refers to the element in the $i^{th}$ row and the $j^{th}$ column and the convex and continuously differentiable function $f(\theta_{ij})$ is defined as}
\begin{equation}
\begin{split}
\; f(\theta_{ij})=\frac{(\theta_{ij}-\theta^{min}_{ij}-\eta_{ij})(\theta_{ij}-\theta^{max}_{ij}+\eta_{ij})}{(\theta^{max}\eta_{ij}-\theta^{min}_{ij}-\eta_{ij})\eta_{ij}},
\end{split}
\end{equation}
\textit{where $\eta_{ij} \in \mathbb{R}^{+}$ is the projection tolerance of $\theta_{ij}$ while $\eta_{ij} < \frac{\theta^{max}_{ij}-\theta^{min}_{ij}}{2}$ and $\eta_{ij}<\theta^{max}_{ij}\; \&\; |\theta^{min}_{ij}|$ hold, and while $\theta^{max}_{ij}>0$ and $\theta^{min}_{ij}<0$ are respectively the upper-bound and the lower-bound of $\theta_{ij}$. Now $Proj(\theta_{ij},\Theta_{ij})$ is a calculable function of our eigen-values discussed in the previous parts, that is, $\mathbb{P}\mathscr{r} \big \lbrace \Lambda \ge \lambda   \big \rbrace$.}
\end{proposition}

\textbf{\textsc{Proof:}} See Appendix \ref{sec:F}.$\; \; \; \blacksquare$

\begin{proposition} \label{P2} \textit{Even if the secrecy rate $ \Lambda$ is not convex and compact, or for more generality, it is not contractible, that can still be relaxed and an equilibrium can be consequently found in relation to the eigenvalues discussed above.}\end{proposition}

\textbf{\textsc{Proof:}} See Appendix \ref{sec:G}.$\; \; \; \blacksquare$

\begin{proposition} \label{P2} \textit{For the case of periodic attacks, an extension over our Markov decision process and with regard to the possibility-theory is satisfied in term of a possibilisitically semi-Markov decision process.}\end{proposition}

\textbf{\textsc{Proof:}} See Appendix \ref{sec:H}.$\; \; \; \blacksquare$

\section{Numerical results}

Initially opening, we have done our simulations w.r.t. the Bernoulli-distributed data-sets using GNU Octave of version $4.2.2$ on Ubuntu $16.04$. 
 
Consider a three-dimensional coordinate network setup consisting of Alice located at $(-50, 0, 0)$, Bob located at $(0, 50 \sqrt3, 0)$ and Eve located at $(0, -d, 0)$ while $d$ is given unkown. Indeed, the distance between Alice and Bob is $D_{AB}=100 \;m$ and the distance between Alice and Eve is $D_{AE}=\sqrt{50^2+d^2} \;m$. The channel matrices are modeled as $H=D_{AB}^{-\epsilon/2}\hat{H}$ and $G=D_{AE}^{-\epsilon/2}\hat{G}$ for respectively Alice-Bob link and Alice-Eve one while $\hat{H}$ and $\hat{G}$ are small-scale Rayleigh fading modeled matrices with i.i.d. complex Gaussian entries with the zero-mean and the variance of $\frac{1}{size}$ where $size$ is the size of the relative matrices, and $\epsilon$ stands for the path-loss exponent $-$ something that is chosen $3$. Consequently, the signal-to-noise-ratio (SNR) values at Bob and Eve are respectively deriven as $SNR_B=\frac{P_{a}}{D_{AB}^{\epsilon}\partial^2 _b}$ and $SNR_E=\frac{P_{a}}{D_{AE}^{\epsilon} \partial^2 _e}$ where $P_a$ stands literally for the transmition power of Alice, while $\partial^2 _b$ and $\partial^2 _e$ are the noise variances at Bob and Eve, respectively.

Table \ref{table2} shows the SOP's convex version vs. $\mathcal{I}\big(  \mathcal{X},\mathcal{A} \big)$ while changing $\rho$ $-$ something that is perfect for the evaluation here. We indeed use $\frac{e^{-\rho^{2}}}{ \mathbb{P}\mathscr{r} \big(     \Lambda  \big)}$, something that can be in connection with another lower-bound from a traditional point of view, that is, $ \mathbb{P}\mathscr{r} \big(  \frac{SNR_B}{SNR_E}   < 2^{R_s}  \big)$ while $R_s$ is our arbitary secrecy rate threshold. A comparison is also made with the Algorithm \ref{foor} given below.

\begin{algorithm}
\caption{\textcolor{black}{A Projection method based algorithm.}}\label{foor}
\begin{algorithmic}
\STATE \textbf{\textsc{Initialisation.}} 

$\;\;$$\;\;$ \textbf{while} $ \mathbb{TRUE}$ \textbf{do}

$\; \; \; \; \; \; \; \; \;$ Compute $  f (\theta_{ij})    $,

$\; \; \; \; \; \; \; \; \;$ Compute $ \nabla f (\theta_{ij}) $,

$ \; \; \; \; \; \; \; \; \;$ \textbf{if} $f(\theta_{ij})>0\; \&\; \Theta_{ij} ( \nabla f (\theta_{ij}) )>0,$

$ \; \; \; \; \; \; \; \; \; \; \;$ Update $Proj(\theta_{ij},\Theta_{ij}) \gets \Theta_{ij}-f (\theta_{ij})\Theta_{ij} $ according to the Eqn. \ref{eq:13}.

$ \; \; \; \; \; \; \; \; \;$ \textbf{elseif}

$ \; \; \; \; \; \; \; \; \; \; \;$ Update $Proj(\theta_{ij},\Theta_{ij}) \gets \Theta_{ij} $ according to the Eqn. \ref{eq:13}.

$\;\;$$\;\;$ \textbf{endif} 

$\;\;$$\;\;$ \textbf{endwhile} 

\textbf{\textsc{Output:}} $Proj(\theta_{ij},\Theta_{ij}) $

\textbf{end}

\end{algorithmic}
\end{algorithm}

\textcolor{black}{Table \ref{table3} and Table \ref{table4} also show the complexity/accuracy comparison for the possible greedy algorithm \ref{f} to find $    \mathop{{\rm \mathbb{M}ax}}\limits_{\Delta = \big \lbrace \lambda_1, \cdots, \lambda_n\big \rbrace  } {\rm \; } \mathbb{E}  \Big \lbrace         \mathbb{V}\mathscr{ol}\big \lbrace  \Lambda   \big \rbrace   \Big \rbrace$ and $ \mathop{{\rm \mathbb{M}ax}}\limits_{\Delta = \big \lbrace \lambda_1, \cdots, \lambda_n\big \rbrace  } {\rm \; }   \mathbb{V}\mathscr{ol}  \Big \lbrace\mathbb{P}\mathscr{r} \big(     \Lambda \ge \lambda \big)     \Big \rbrace $. As obvious, it is proven that our derived bound is more acceptable.}

\begin{table*}[h]
\begin{center}
\captionof{table}{{Simulations: SOP's convex version vs. $\mathcal{I}\big(  \mathcal{X},\mathcal{A} \big)$ while changing $\rho$.}}
 \label{table2} 
\begin{tabular}{p{1.5cm}|p{3.3cm} |p{1.5cm}|p{3.6cm}|p{1.5cm}|p{3.6cm}}
 \hline      
 \hline
$\mathcal{I}\big(  \mathcal{X},\mathcal{A} \big)$        &  SOP's convex version  & $\mathcal{I}\big(  \mathcal{X},\mathcal{A} \big)$  & SOP's convex version         &  $\mathcal{I}\big(  \mathcal{X},\mathcal{A} \big)$  & SOP's convex version   \\
\hline
  $0$   &                   $Our=0.005, \rho=0.1$   &           $0.5$    &     $Our=0.0026, \rho=0.1$ &           $1$    &    $Our=0.0001, \rho=0.1$          \\
    $0$   &                   $Our=0.005, \rho=0.2$      &           $0.5$    &    $Our=0.0027, \rho=0.2$ &           $1$    &     $Our=0.0002, \rho=0.2$          \\
 \hline
$0$   &                   $Proj=0.005, \rho=0.1$   &           $0.5$    &     $Proj=0.0027, \rho=0.1$ &           $1$    &    $Proj=0.00011, \rho=0.1$          \\
    $0$   &                   $Proj=0.005, \rho=0.2$      &           $0.5$    &    $Proj=0.00274, \rho=0.2$ &           $1$    &     $Proj=0.0002, \rho=0.2$          \\
\hline
 \hline
\end{tabular}
\end{center}
\end{table*}

\begin{table*}[h]
\begin{center}
\captionof{table}{{\textcolor{black}{Simulations: Complexity vs. Iterations derived from $ \Delta=\mathop{{\rm \mathbb{M}ax}}\limits_{\Delta = \big \lbrace \lambda_1, \cdots, \lambda_n\big \rbrace  } {\rm \; }   \mathbb{V}\mathscr{ol}  \Big \lbrace\mathbb{P}\mathscr{r} \big(     \Lambda \ge \lambda \big)     \Big \rbrace $  divided by the derived one from $   \Delta= \mathop{{\rm \mathbb{M}ax}}\limits_{\Delta = \big \lbrace \lambda_1, \cdots, \lambda_n\big \rbrace  } {\rm \; } \mathbb{E}  \Big \lbrace         \mathbb{V}\mathscr{ol}\big \lbrace  \Lambda   \big \rbrace   \Big \rbrace$}.}}
 \label{table3} 
\begin{tabular}{p{1.5cm}|p{2.3cm} |p{1.5cm}|p{2.3cm}|p{1.5cm}|p{2.3cm}}
 \hline      
 \hline
Iterations  & Complexity        &  Iterations  & Complexity & Iterations  & Complexity \\
\hline
  $0$   &                   $0.99$   &           $50$    &     $0.94$ &           $100$    &    $0.93$     \\    
\hline
 \hline
\end{tabular}
\end{center}
\end{table*}

\begin{table*}[h]
\begin{center}
\captionof{table}{{\textcolor{black}{Simulations: Accuracy vs. Iterations derived from $ \Delta=\mathop{{\rm \mathbb{M}ax}}\limits_{\Delta = \big \lbrace \lambda_1, \cdots, \lambda_n\big \rbrace  } {\rm \; }   \mathbb{V}\mathscr{ol}  \Big \lbrace\mathbb{P}\mathscr{r} \big(     \Lambda \ge \lambda \big)     \Big \rbrace $  divided by the derived one from $ \Delta=   \mathop{{\rm \mathbb{M}ax}}\limits_{\Delta = \big \lbrace \lambda_1, \cdots, \lambda_n\big \rbrace  } {\rm \; } \mathbb{E}  \Big \lbrace         \mathbb{V}\mathscr{ol}\big \lbrace  \Lambda   \big \rbrace   \Big \rbrace$}.}}
 \label{table4} 
\begin{tabular}{p{1.5cm}|p{2.3cm} |p{1.5cm}|p{2.3cm}|p{1.5cm}|p{2.3cm}}
 \hline      
 \hline
Iterations  & Accuracy        &  Iterations  & Accuracy & Iterations  &Accuracy \\
\hline
  $0$   &                   $1.01$   &           $50$    &     $1.1$ &           $100$    &    $1.12$     \\    
\hline
 \hline
\end{tabular}
\end{center}
\end{table*}

\begin{figure}[t]
\centering
\subfloat[Average optimal policy $\pi_{t}(a|s)$]{\includegraphics[trim={{17 mm} {64 mm} {21 mm} {74mm}},clip,scale=0.4]{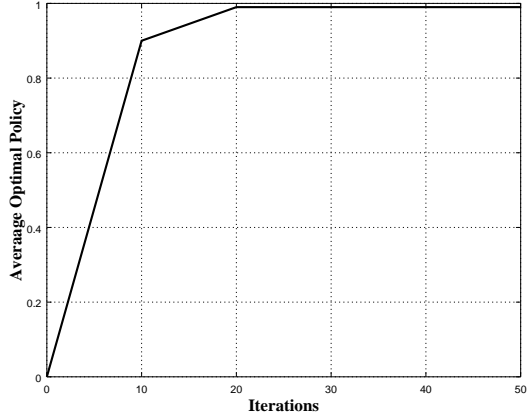}} \\
\subfloat[Average reward $\mathcal{R} \big(   \mathcal{S}_{t},\mathcal{A}_{t}  \big)$]{\includegraphics[trim={{17 mm} {64 mm} {21 mm} {74mm}},clip,scale=0.4]{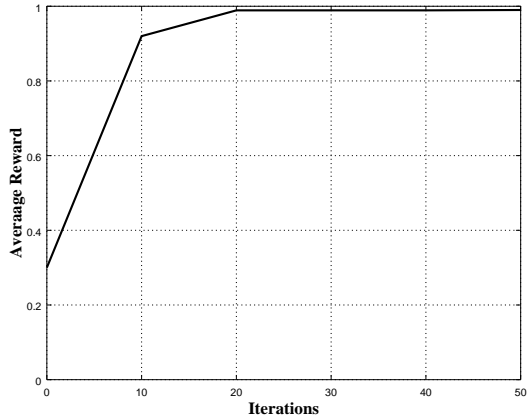}} \\
\subfloat[Error for Q-function $\mathcal{Q}_{t}(s,a)$]{\includegraphics[trim={{17 mm} {64 mm} {21 mm} {74mm}},clip,scale=0.4]{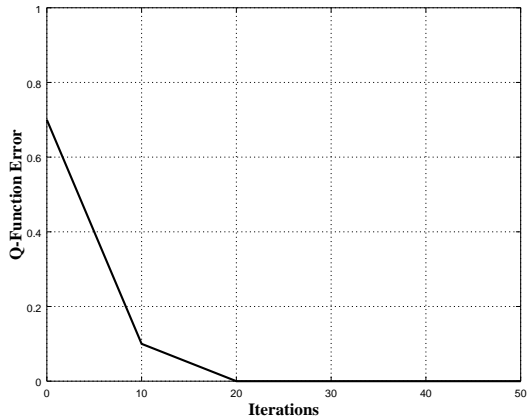}} 
\caption{Reward, policy and Q-function error versus the iteration regime, while $|\mathcal{S}|=  \mathscr{U}_0+\mathscr{V}_0=5+7=12$, $|\mathcal{A}|=max \{  \mathscr{U}_0, \mathscr{V}_0  \}=7$, $\mathcal{V}_{ensembel}(\zeta)=\zeta^2$.} \label{F11}
\label{fig:EcUND} 
\end{figure}

Finally speaking, in Fig. \ref{F11}, and the subfigures \textit{a}, \textit{b} and \textit{c} the average reward $\mathcal{R} \big(   \mathcal{S}_{t},\mathcal{A}_{t}  \big)$, the error in relation to the Q-function $\mathcal{Q}_{t}(s,a)$, i.e., $\frac{\mathcal{Q}_{t}-\mathcal{Q}^{\star}}{\mathcal{Q}^{\star}}$ and the average optimal policy $\pi_{t}(a|s)$ $-$ while $(\cdot)^{\star}$ stands for the optimum-value $-$ are respectively depicted versus the iteration regime while $|\mathcal{S}|=  \mathscr{U}_0+\mathscr{V}_0=5+7=12$, $|\mathcal{A}|=max \{  \mathscr{U}_0, \mathscr{V}_0  \}=7$, $\mathcal{V}_{ensembel}(\zeta)=\zeta^2$ are basically selected.

\section{conclusion}
 A new bound and the relating interpretations over the concave version of the SOP maximisation problem were fundamentally explored in this paper. We technically considered a Riemannian mani-fold for the SOP's concave version and a volume for it. Towards such end, some highly professional and insightful principles such as Keyhole contour, Finsler’s lemma, the generalised Brunn-Minkowski inequality etc were used. In order to find the optimal policy in relation to the eigenvalue distributions, a novel Markov decision process based reinforcement learning algorithm was also essentially proposed $-$ something that was subsequently extended to a possibilisitically semi-Markov decision process for the case of periodic attacks and with regard to the possibility-theory.

\appendices
\section{Proof of Lemma 1}
\label{sec:A}
The proof is performed according to the Taylor expansion of
\begin{equation}
\begin{split}
\; e^{\mathscr{t} \Lambda} \textcolor{black}{\; =\; } 1+\mathscr{t} \Lambda \textcolor{black}{\;   +\;  \mathscr{O} \{ \cdot \}   },
\end{split}
\end{equation}
\textcolor{black}{while $\mathscr{O} \{ \cdot \} $ is the \textit{Big-O notation}}. Now, by applying an expected-value operand, we consequently reach out 
\begin{equation}
\begin{split}
\; \mathbb{E} \big \lbrace \Lambda  \big \rbrace \textcolor{black}{\; \le\; } \frac{\mathbb{E} \big \lbrace  e^ { \mathscr{t} \Lambda  } \big \rbrace-1}{\mathscr{t}}.
\end{split}
\end{equation}

The proof is now completed.$\; \; \; \blacksquare$

\section{Proof of Lemma 4}
\label{sec:B}
The cumulative distribution function (CDF) 
\begin{equation}
\begin{split}
\; \mathscr{F}_{\Lambda} (\lambda)=\mathbb{P}\mathscr{r} \big(     \Lambda \le \lambda \big)\;\;\;\;\;\;\;\;\;\\=1-\mathbb{P}\mathscr{r} \big(     \Lambda \ge \lambda \big)\;\;\;\\=1-e^ { -\mathscr{t} \Lambda} \mu_{\Lambda}(\mathscr{t}) \;\;,
\end{split}
\end{equation}
holds while $\mu_{\Lambda}(\mathscr{t})$ is the moment-generating function (MGF), so, we have
\begin{equation}
\begin{split}
\; \mathbb{E} \big \lbrace  e^ { \mathscr{t} \Lambda  } \big \rbrace=e^ { \mathscr{t} \lambda} \mathbb{P}\mathscr{r} \big(     \Lambda \ge \lambda \big),
\end{split}
\end{equation}
holds.

The proof is now completed.$\; \; \; \blacksquare$

\section{Proof of Lemma 5}
\label{sec:C}
Let us assume that we have the optimisation problem of $           \mathop{{\rm \mathbb{M}ax}}\limits_{\Delta = \big \lbrace \lambda_1, \cdots, \lambda_n\big \rbrace  } {\rm \; } \mathbb{E}  \Big \lbrace         \mathbb{V}\mathscr{ol}\big \lbrace  \Lambda   \big \rbrace   \Big \rbrace $, something that is equivalent to the maximisation over its supremum as in
\begin{equation}
\begin{split}
\;   \mathop{{\rm \mathbb{M}ax}}\limits_{\Delta = \big \lbrace \lambda_1, \cdots, \lambda_n\big \rbrace  } {\rm \; } \mathbb{V}\mathscr{ol}  \Big \lbrace              \frac{\mathbb{E} \big \lbrace e^ { \mathscr{t} \Lambda  } \big \rbrace  -1}{\mathscr{t}}     \Big \rbrace,
\end{split}
\end{equation}
or with the aid of \textit{Lemma 4} $-$ the Eqn. \ref{eq:5} $-$ , as in 
\begin{equation}
\begin{split}
\;     \mathop{{\rm \mathbb{M}ax}}\limits_{\Delta = \big \lbrace \lambda_1, \cdots, \lambda_n\big \rbrace  } {\rm \; }  \mathbb{V}\mathscr{ol}  \Big \lbrace              \frac{e^ { \mathscr{t} \lambda} \mathbb{P}\mathscr{r} \big(     \Lambda \ge \lambda \big)  -1}{\mathscr{t}}     \Big \rbrace ,
\end{split}
\end{equation}
or
\begin{equation}
\begin{split}
\;        \mathop{{\rm \mathbb{M}ax}}\limits_{\Delta = \big \lbrace \lambda_1, \cdots, \lambda_n\big \rbrace  } {\rm \; }  \mathbb{V}\mathscr{ol}  \Big \lbrace              e^ { \mathscr{t} \lambda} \mathbb{P}\mathscr{r} \big(     \Lambda \ge \lambda \big)     \Big \rbrace ,
\end{split}
\end{equation}
or finally 
\begin{equation}
\begin{split}
\;  \mathop{{\rm \mathbb{M}ax}}\limits_{\Delta = \big \lbrace \lambda_1, \cdots, \lambda_n\big \rbrace  } {\rm \; }  \mathbb{V}\mathscr{ol}  \Big \lbrace              e^ { \mathscr{t} \lambda}  \Big \rbrace \mathbb{V}\mathscr{ol}  \Big \lbrace\mathbb{P}\mathscr{r} \big(     \Lambda \ge \lambda \big)     \Big \rbrace .
\end{split}
\end{equation}

The proof is now completed.$\; \; \; \blacksquare$

\begin{algorithm*}
\caption{\textcolor{black}{A two-time-scale natural actor-critic $\epsilon-$greedy algorithm.}}\label{fooor}
\begin{algorithmic}
\STATE \textbf{\textsc{Input.}} \begin{itemize}
\item $\mathcal{S}_0 \sim$ \textit{arbitary}; $\;\;\;\;\; //$ Initial state
\item $\mathcal{A}_0 \sim \pi_o (\cdot|s_0)$; $\;\;\;\;\; //$ Initial action
\item $ \pi_o (a|s)=\hat{\pi}_o (a|s)=\frac{1}{|\mathcal{A}|}$; $\;\;\;\;\; //$ Initial ploicy to be learnt, as the probability distribution of taking the relative action at the given state
\item $T>0$; $\;\;\;\;\; //$ Iteration number
\item $\alpha_t>0$; $\;\;\;\;\; //$ Step size
\item $\beta_t>0$; $\;\;\;\;\; //$ Step size
\item $\epsilon_t>0$;  $\;\;\;\;\; //$ Greedy factor
\item $\psi>0$;  $\;\;\;\;\; //$ Discount factor
\item $\mathcal{Q}_0(s,a) \in \mathbb{R}^{|\mathcal{S}||\mathcal{A}|}$. $\;\;\;\;\; //$ Initial Q-function corresponding to the initial policy $\pi_o (a|s)$
\end{itemize}

$\;\;$$\;\;$ \textbf{for} $t=0, 1, \cdots, T$ \textbf{do} 

$\; \; \; \; \; \; \; \; \;$ Sample $\mathcal{S}_{t+1}\sim \mathcal{P}\big( \cdot|\mathcal{S}_{t},\mathcal{A}_{t}    \big)$; $\;\;\;\;\; //$ According to the set of the transition probability matrices $ \mathcal{P}$

$\; \; \; \; \; \; \; \; \;$ Sample $\mathcal{A}_{t+1}\sim \hat{\pi}_t\big( \cdot|\mathcal{S}_{t+1}    \big)$;

$\; \; \; \; \; \; \; \; \;$ $\alpha_t(s,a) \gets \alpha_t \mathbbm{1} \big \lbrace   \mathcal{S}_{t}=s,\mathcal{A}_{t}=a, \forall s,a   \big \rbrace$;

$\; \; \; \; \; \; \; \; \;$ $\mathcal{Q}_{t+1}(s,a) \gets \mathcal{Q}_{t}(s,a)+\alpha_t(s,a)\Big[  \mathcal{R} \big(   \mathcal{S}_{t},\mathcal{A}_{t}  \big)  +\psi  \mathcal{Q}_{t} \big(   \mathcal{S}_{t+1},\mathcal{A}_{t+1}  \big)  -\mathcal{Q}_{t}\big(   \mathcal{S}_{t},\mathcal{A}_{t}  \big)    \Big], \forall s,a$;  $\;\;\;\;\; //$ With regard to the \textit{Reward-function} $\mathcal{R} (\cdot,\cdot) \propto f\Big(   \frac{1}{\mathcal{P} \big(\zeta^{(\cdot)}_{max},\zeta^{(\cdot)}_{min} \big)}    \Big)$ while $\mathcal{P} \big(\zeta^{(\cdot)}_{max},\zeta^{(\cdot)}_{min} \big)$ to be chosen

$\; \; \; \; \; \; \; \; \;$ $\pi_{t+1}(a|s) \gets \pi_{t}(a|s) \frac{exp \big (  \beta_t\mathcal{Q}_{t+1}(s,a)   \big)}{\sum_{a^{\prime}}\pi_{t}(a^{\prime}|s)exp \big (  \beta_t\mathcal{Q}_{t+1}(s,a^{\prime})   \big)}, \forall s,a$;

$\; \; \; \; \; \; \; \; \;$ $\hat{\pi}_{t+1}\gets \frac{\epsilon_t}{|\mathcal{A}|}+(1-\epsilon_t)\pi_{t+1}$.

$\;\;$$\;\;$ \textbf{endfor} 

Sample $\hat{T}$ from $\{ 0, 1, \cdots, T\}$ by distribution $\mathcal{P} \big(  \hat{T}=i  \big)=\frac{\beta_i}{\sum_{j=0}^{T}\beta_j}$.

\textbf{\textsc{Output:}} $\hat{\pi}_{\hat{T}} $

\textbf{end}

\end{algorithmic}
\end{algorithm*}

\section{Proof of Proposition 1}
\label{sec:D}
The proof is provided here in terms of the following solution. 

Where $K$ is a constant scaling factor, one can re-write the polynomial as\footnote{See e.g. \cite{2}.}
\begin{equation}
\begin{split}
\;  \mathscr{P}(z)=K  \prod_{i=1}^{n} (z-\zeta_i), 
\end{split}
\end{equation}
while $\zeta_i, i \in \{1, \cdots, n\}$ stands literally for the $i-$th eigen-value.

Now, recall the term $\mathbb{V}\mathscr{ol}  \Big \lbrace\mathbb{P}\mathscr{r} \big(     \Lambda \ge \lambda \big)     \Big \rbrace$ versus $\int \mathbb{V}\mathscr{ol}  \Big \lbrace\mathbb{P}\mathscr{r} \big(     \Lambda \ge \lambda \big)     \Big \rbrace$. By differentiating $\mathscr{P}(z)$ with respect to $z$ as $\mathscr{P}^{\prime}(z)$, $\frac{\mathscr{P}^{\prime}(z)}{\mathscr{P}(z)}$ is obtained as
\begin{equation}
\begin{split}
\;  \frac{\mathscr{P}^{\prime}(z)}{\mathscr{P}(z)}=  \sum_{i=1}^{n} \frac{1}{z-\zeta_i}.
\end{split}
\end{equation}

For the above equation, where $j =\sqrt {-1}$ is the imaginary unit, $\mathscr{L} \supseteq \mathbb{V}\mathscr{ol}  \Big \lbrace\mathbb{P}\mathscr{r} \big(     \Lambda \ge \lambda \big)     \Big \rbrace$ is a closed anti-clockwise curve on the complex plane, and $\mathscr{C} \supseteq \int \mathbb{V}\mathscr{ol}  \Big \lbrace\mathbb{P}\mathscr{r} \big(     \Lambda \ge \lambda \big)     \Big \rbrace$ is the region enclosed by $\mathscr{L}$, it is achieved as\footnote{See e.g. \cite{3}.}
\begin{equation}
\begin{split}
\;  \oint_{\mathscr{L} }  \sum_{i=1}^{n} \frac{1}{z-\zeta_i}dz=\begin{cases}
2 \pi j, \; \; \; \; \mathsf{if} \; \zeta_i \in \mathscr{C} ,\\
0, \; \; \; \; \; \; \; \; \mathsf{if} \;  \zeta_i \notin \mathscr{C} ,
\end{cases}
\end{split}
\end{equation}
accoring to which one can say that the number of the eigen-values in the region $\mathscr{C} $ is
\begin{equation}
\begin{split}
\;  N=\frac{1}{2 \pi j}\oint_{\mathscr{L} } \frac{\mathscr{P}^{\prime}(z)}{\mathscr{P}(z)}dz\;\;\;\;\;\;  \\=\frac{1}{2 \pi j}\sum_{i=1}^{n} \oint_{\mathscr{L} }  \frac{1}{z-\zeta_i}dz.
\end{split}
\end{equation}

On the other hand, $\mathscr{P}^{\prime}(z)$ is obtained as \cite{4}\footnote{Page 8, eqn. 46.}
\begin{equation}
\begin{split}
\;  \mathscr{P}^{\prime}(z)=det(\mathcal{B}-z\mathcal{A}) Tr \Big[(\mathcal{B}-z\mathcal{A})^{-1} \underbrace{\frac{\partial (\mathcal{B}-z\mathcal{A})}{\partial z}}_{-\mathcal{A}} \Big],
\end{split}
\end{equation}
while $Tr [\cdot]$ stands for the trace of the matrix, something that is equivalent to
\begin{equation}
\begin{split}
\;  \mathscr{P}^{\prime}(z)=\mathscr{P}(z) Tr \Big[(\mathcal{B}-z\mathcal{A})^{-1} (-\mathcal{A}) \Big],
\end{split}
\end{equation}
according to which one can say 
\begin{equation}
\begin{split}
\;  N=\frac{1}{2 \pi j}\oint_{\mathscr{L} }   Tr \Big[(\mathcal{B}-z\mathcal{A})^{-1} (-\mathcal{A}) \Big]     dz.
\end{split}
\end{equation}

The last integral, i.e., the equation appeared above can be efficiently solved by some digitised methods such as the \textit{Rayleigh-Ritz method} \cite{r, rr, rrr}.

In order to conclude the proof, let us ultimately go over the essential relevance between the number of eigen-values and $\mathbb{V}\mathscr{ol}  \Big \lbrace\mathbb{P}\mathscr{r} \big(     \Lambda \ge \lambda \big)     \Big \rbrace$ in the context of the following lemma. 

\begin{lemma} \label{P2} \textit{The number of eigen-values discussed above relies fundamentally upon $\mathbb{P}\mathscr{r} \big(     \Lambda  \big)$.}\end{lemma} 

\textit{Proof.} In relation to the term $ \mathbb{V}\mathscr{ol}  \Big \lbrace\mathbb{P}\mathscr{r} \big(     \Lambda \ge \lambda \big)     \Big \rbrace$, we get in hands
\begin{equation}
\begin{split}
\;   \mathbb{P}\mathscr{r} \Big(\big(     \Lambda \ge \lambda \big) \ge \rho \Big)   \le 1-\frac{e^{-\rho^{2}}}{ \mathbb{P}\mathscr{r} \big(     \Lambda  \big)}
\end{split}
\end{equation}
according to the \textit{Talagrand's Concentration inequality}\footnote{See e.g. \cite{7} to understand what it is: It says that the complement of the given random variable in a bounded probability closure is emphatically upperbounded.}, while $\rho$ is an arbitary threshold. This means that $\mathbb{V}\mathscr{ol}  \Big \lbrace\mathbb{P}\mathscr{r} \big(     \Lambda \ge \lambda \big)     \Big \rbrace$, $\mathscr{C}$ and $\mathscr{L}$ are functions of $\Big(\rho;  \mathbb{P}\mathscr{r} \big(     \Lambda  \big)\Big)$ $-$ something that proves \textit{Lemma 8}.

\textsc{\textbf{Remark 4.}} \textit{The accuracy of evaluating the eigen-values expressed here can be fully able to be controlled by $\rho$.}

The proof is now completed.$\; \; \; \blacksquare$

\section{Proof of Proposition 2}
\label{sec:E}
The sketch of the proof is given here which is similar to \cite{vvvv}. 

We know \cite{vvvv}
\begin{equation}
\begin{split}
\;  \mathbb{P}\mathscr{r} \big( L \leq (1-\tau_2)\tau_1  \big ) \;\;\;\;\;\;\;\;\;\;\;\;\;\;\;\;\;\;\;\;\;\;\;\;\;\;\;\;\;\;\;\;\;\;\;\;\;\;\;\; \\    \leq   exp  \big (    \frac{k}{2} (1-(1-\tau_2) + ln (1-\tau_2)  \big ),
\end{split}
\end{equation}
which can be re-casted to 
\begin{equation}
\begin{split}
\;  \mathbb{P}\mathscr{r} \big( L \leq (1-\tau_2)\tau_1  \big ) \;\;\;\;\;\;\;\;\;\;\;\;\;\;\;\;\;\;\;\;\;\;\;\;\;\;\;\;\;\;\;\;\;\;\;\;\;\;\;\; \\    \leq exp \bigg (  \frac{k}{2} \Big ( \tau_2-\big ( \tau_2+\frac{\tau^2_2}{2} \big)\Big)\bigg),
\end{split}
\end{equation}
since $ln(1-x) \leq -x-\frac{x^2}{2}, \forall \; 0 \leq x <1$ holds $-$ something that can conclude the proof.$\; \; \; \blacksquare$

\section{Proof of Proposition 3}
\label{sec:F}
The proof is given as the following. 

We know \cite{cvcv}
\begin{equation}
\begin{split}
\;  tr \Big(( \theta^{T}-\theta^{*T}) \big(-\Theta+Proj(\theta,\Theta) \big) \Big)<0,
\end{split}
\end{equation}
holds for $\theta^{*} \in [\theta^{min}_{ij}+\eta_{ij}, \theta^{max}_{ij}-\eta_{ij}]$ while the trace operator $tr(\cdot)$ is a function of $-$ sum of $-$ the eigen-values related to $\theta$, that is, $\Lambda$ in our scheme and analysis.


We furthermore know that \textit{Sizes of random projections of sets}, i.e., Thereom $7.7.1$ in \cite{Vershynin} may help us to prove that if we have a bounded set $\theta \in \mathbb{R}^{r_1}, r_1 <r$ while $r$ was defined in Proposition 2, with a projected set $\Theta \in \mathbb{R}^{r_2}, r_2 <r$, with a probability of at least $1-2e^{-r_2}$ we have 
\begin{equation}
\begin{split}
\;  diam\big(  \Theta\theta  \big)\le C_0 \Big(  w_s \big( \theta  \big)+\sqrt{\frac{r_2}{r_1}} diam\big(  \theta  \big)\Big),
\end{split}
\end{equation}
while $C_0$ is a constant, $diam(\cdot)$ stands for the diameter, and $w_s(\theta)$ denotes the \textit{Gaussian width} as $\mathbb{E} \mathop{{\rm \mathbb{S}up}}\limits_{x \in  \theta} {\rm \; }\langle  x, \mathscr{g} \rangle, \mathscr{g} \sim \mathscr{N}(0,I_{r_1})$.


The proof is now completed.$\; \; \; \blacksquare$

\section{Proof of Proposition 4}
\label{sec:G}
Let us start the proof with the \textit{Gauss-Bonnet-Theorem}\footnote{See e.g. \cite{Julia1, Julia2} to understand what it is.}. It says that for a manifold $\mathcal{M}$ with the boundary $\partial \mathcal{M}$, with the \textit{Euler characterisitcs} $\mathcal{X}\big(\mathcal{M}\big)$ and the \textit{Gaussian Curvature}\footnote{See e.g. \cite{GIL, KHAYEH20, LEE1, LEE2} to understand what it is.} $\mathcal{K}$ and the \textit{Geodesic Curvature}\footnote{See e.g. \cite{GIL, KHAYEH20, LEE1, LEE2} to understand what it is. } $\mathcal{K}_g$ relating to $\partial \mathcal{M}$, the following is satidfied
\begin{equation}
\begin{split}
\;  \int_{\mathcal{M}} \mathcal{K}d \mathcal{S}_{area} + \int_{\partial \mathcal{M}} \mathcal{K}_g ds =2 \pi \mathcal{X}\big(\mathcal{M}\big),
\end{split}
\end{equation}
while $\mathcal{S}_{area}$ stands theoretically for the area of $\mathcal{M}$ and $s \subset \mathcal{S}_{area}$.

Now, we initially see that the contractibility radius as well as the equilibrium we are supposed to go over rely deeply upon the principal curvatures, i.e., the eigen-vectors. 

Additionally, \textit{Pu-1952 inquality}\footnote{\cite{KISEH}.} says that the \textit{Systol} of a manifold $\mathcal{M}$ as $\mathcal{SYS}\big (  \mathcal{M} \big)$ as the least lenght\footnote{\cite{KOOOSE}.} of a non-contractible loop of the homeomorphic manifold $\mathcal{M}$ $-$ to the real projective plan $-$ , i.e., the lowerbound of the lenghts of non-contractible closed curves over $\mathcal{M}$\footnote{That is, $\stackrel{def}{=}inf \big \lbrace   \mathscr{l}(c) | \;c:$ non-contractible closed curves    $\big \rbrace$ from a mathematical point of view, while $ \mathscr{l}(c)$ denotes the lenght of $c$.} satisfies 
\begin{equation}
\begin{split}
\;  \mathcal{S}_{area} \ge \frac{2}{\pi}\mathcal{SYS}^2 \big (  \mathcal{M} \big),
\end{split}
\end{equation}
while the equlity holds\footnote{\textit{Minding's} theorem.} for the \textit{constant Gaussian curvatures}, i.e., when $\mathcal{M}$ is \textit{locally isometric}. Or correspondingly\footnote{See e.g. \cite{JIMI1, JIMI2}.}, 
\begin{equation}
\begin{split}
\;  \mathcal{C}_n \Big(\mathbb{V}\mathscr{ol}  \big(\mathcal{M}\big)\Big)^{\frac{1}{n}} \ge \mathcal{SYS} \big (  \mathcal{M} \big), \exists \mathcal{C}_n \in \mathbb{R}^{n}.
\end{split}
\end{equation}

Thus, it has so far been proven that, in order to work on the contractibility radius as well as the equilibrium discussed above, it is necessary and sufficient for us to only focus on the eigen-vectors.

Now, in every kind of manifold and space, there may exist multiple maximum-eigenvalues or/and minimum-eigenvalues, for example, a hemisphere has $3$ maximum-eigenvalues and only $1$ minimum-eigenvalue. However, the distribution of the eigenvlaues may be totally different in every case\footnote{See e.g. \cite{Juli1, Juli2, Juli3, Juli4}.}. Thus, there may exist a Markov-Decision-Process $-$ something that technically enforces us to propose the following Reinforcement-learning based algorithm to find the perfect policy according to the eigenvalues' distributions.

\begin{figure}[t]
\centering
\subfloat{\includegraphics[trim={{47mm} {58 mm} {23mm} {44mm}},clip,scale=0.4]{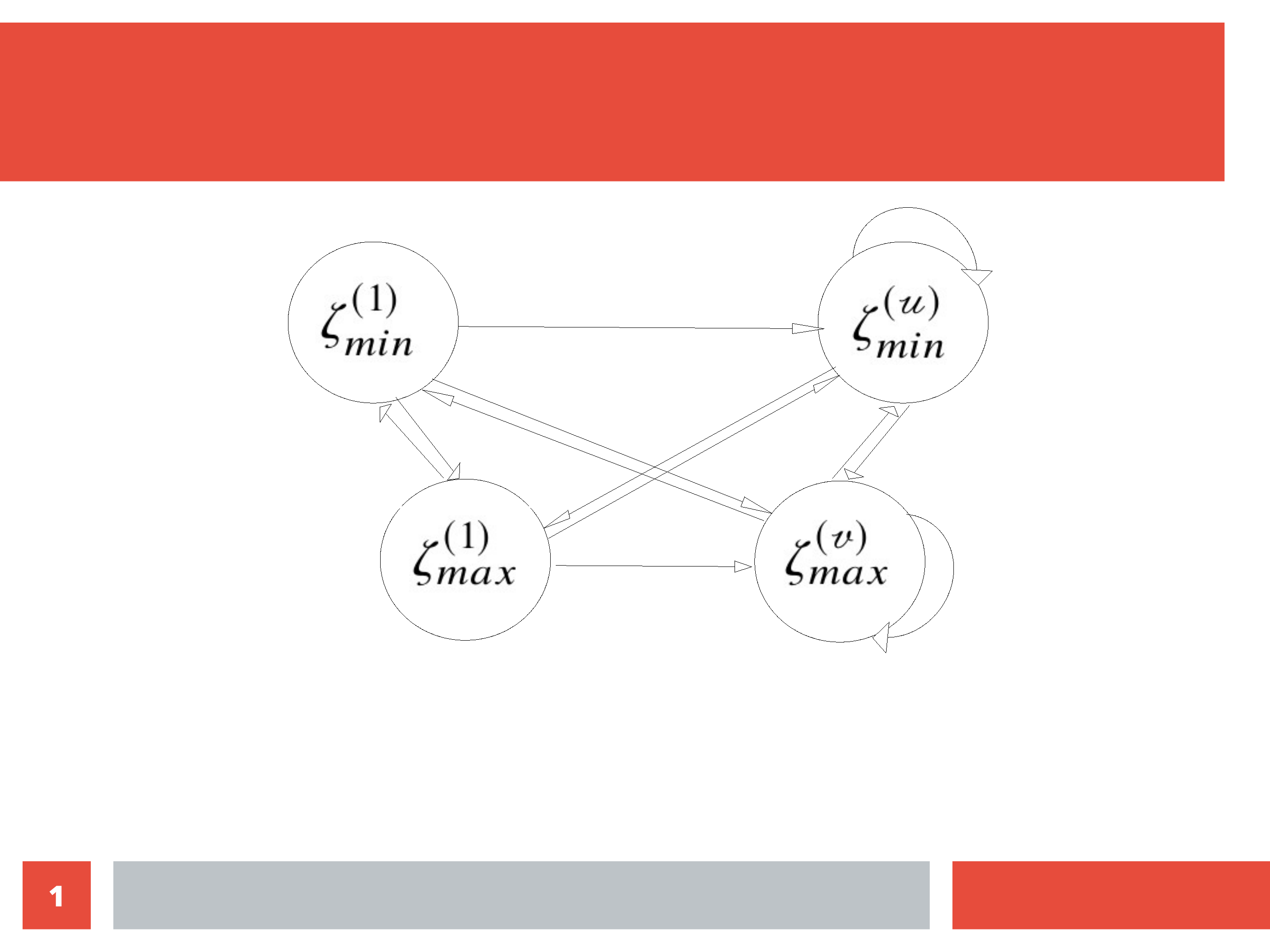}}
\caption{The Markov model in relation to our scheme and $\big \lbrace \zeta^{(1)}_{max}, \cdots, \zeta^{(\mathscr{v})}_{max} \big \rbrace$ as well as the $\big \lbrace\zeta^{(1)}_{min}, \cdots, \zeta^{(\mathscr{u})}_{min}\big \rbrace$
. }
\label{F3}
\end{figure}

In relation to Algorithm \ref{fooor}\footnote{In order to understand a Markov decision process reinforcement learning based algorithm, see e.g. \cite{Juli5}.}, the reward function $\mathcal{R}(\cdot,\cdot)$ plays a vital role. Regarding the fact that we aim at finding an equilibrium as discussed above, and due to the fact that in equilibria, the maximum-eigenvalues and the minimum ones tend to get closer to each other\footnote{See e.g. \cite{Juli1, Juli2, Juli3, Juli4}.} as much as possible, one may select the reward function $\mathcal{R}(\cdot,\cdot)$ as $f\Big(   \frac{1}{\mathcal{P} \big(\zeta^{(\cdot)}_{max},\zeta^{(\cdot)}_{min} \big)}    \Big)$ while $\mathcal{P} \big(\zeta^{(\cdot)}_{max},\zeta^{(\cdot)}_{min} \big)$ stands fundamentally for the joint probability distribution for  the maximum-eigenvalues and the minimum ones $-$ something that is\footnote{See e.g. \cite{Juli3, Juli6, Juli7}.} writen as $\mathcal{P} \big(\zeta^{(\cdot)}_{max},\zeta^{(\cdot)}_{min} \big)\stackrel{def}{=}\prod_{\mathscr{u}}\prod_{\mathscr{v}}|\zeta^{(\mathscr{v})}_{max}-\zeta^{(\mathscr{u})}_{min}|^{\beta}\prod_{\mathscr{u}+\mathscr{v}}e^{-\beta\frac{ \mathscr{U}_0+ \mathscr{V}_0}{2}}\mathcal{V}_{ensembel}(\zeta), \mathscr{u} \in \{  1,\cdots, \mathscr{U}_0  \}, \mathscr{v} \in \{  1,\cdots, \mathscr{V}_0  \}$ while $\beta$ is the \textit{Dyson index} and $\mathcal{V}_{ensembel}(\zeta)$ is theoretically the \textit{variant ensemble} which is e.g. for Gaussian case $\zeta^2$.

\textsc{\textbf{Corollary 2 $-$ Example 1.}} \textit{In case of $\mathcal{X}\big(\mathcal{M}\big)=1$, the principal eigenvalues and the term $\int_{\mathcal{M}} \mathcal{K}d \mathcal{S}_{area} + \int_{\partial \mathcal{M}} \mathcal{K}_g ds $ expressed before have a structure such as $\oint\frac{2}{1+x^2}dx$ $-$ something that is well-routine for information-theoretic schemes such as the dirty-paper-coding-principle. According to what we have gone over in Definition 2, this case guarantees the convexity over the secrecy rate.}

\textsc{\textbf{Corollary 3 $-$ Example 2.}} \textit{In case of $\mathcal{X}\big(\mathcal{M}\big)=0$ e.g. for Torus or Kelin-Bottle, regarding the inequality $ \mathcal{S}_{area} \ge \frac{2}{\pi}\mathcal{SYS}^2 \big (  \mathcal{M} \big)$, it is proven that one should send little amount of information in the sense that a less amount of information leaked by Eve can be guaranteed, aimed at reducing the amount of non-contractibility and tits relative radius. This case and interpretation can be proven as follows according to the Excision-Theorem\footnote{See e.g. \cite{ALEX0} to understand what it says.}. This theorem says that if $\mathcal{M}_0 \subset \mathcal{M}_1 \subset \mathcal{M}$, we say $\mathcal{M}_0$ can be excised if the inclusion map $\big(\mathcal{M} \setminus  \mathcal{M}_0,\mathcal{M}_1 \setminus \mathcal{M}_0\big)$ has an isomorphism relationship\footnote{Duality.} with $\big(  \mathcal{M},\mathcal{M}_1 \big)$. This kind of interpretation can also be proven by the concept of symplectic capacity as described in the following remark.}

\textsc{\textbf{Remark 5 $-$ Symplectic capacity\footnote{See e.g. \cite{JIMI1, JIMI2}.}.}} \textit{The principle of Symplectic capacity falls in finding a contractible periodic orbit whose period bounds the Hofer-Zehnder capacity on the energy level which is related to the cylindrical capacity as follows. It says for $\mathcal{M}_0 \subset \mathcal{M}_1 \subset \mathcal{M}$, the following capacity inequality holds while $\mathcal{C}_{apacity}$ denotes the capacity: $\frac{\mathcal{C}_{apacity}\big(   \mathcal{M}_1\big)}{\mathcal{C}_{apacity}\big(   \mathcal{M}_0\big)} \leq \Big(  \frac{\mathbb{V}\mathscr{ol} \big( \mathcal{M}\big)}{\mathbb{V}\mathscr{ol} \big( \mathcal{M}_0\big)} \Big)^{\frac{1}{n}}$.}

\textsc{\textbf{Corollary 4 $-$ Example 3.}} \textit{In case of complex or/and imaginary values\footnote{See e.g. \cite{ALEX1, ALEX2, ALEX3} to understand what they technically are. See also \textit{Caldero-Chapoton function}.} such as $\mathcal{X}\big(\mathcal{M}\big)=i$, the principal eigenvalues and the term $\int_{\mathcal{M}} \mathcal{K}d \mathcal{S}_{area} + \int_{\partial \mathcal{M}} \mathcal{K}_g ds $ expressed before have a structure such as $\oint\frac{1}{x}dx$ $-$ something that may result in creation of bifurcations in eigenvalues.}

\textsc{\textbf{Remark 6 $-$ Conformal equivalence for curvatures\footnote{See e.g. \cite{Shyla}.}.}} \textit{Two metrics $\mathscr{g}_0$ and $\mathscr{g}_{\phi}$ are conformally equivalent if $\mathscr{g}_{\phi}=e^{2\phi}\mathscr{g}_0$ holds while $e^{2\phi}$ is called the conformal factor. Now, the following is satisfied for the relative curvatures \cite{Shyla}: $\mathcal{K}_{\phi}=e^{2\phi} \big(\mathcal{K}_0-\Delta\phi \big)$ while $\Delta\phi$ is the Laplacian on the relative surface.}

\textsc{\textbf{Remark 7 $-$ Davis-Kahan-Theorem\footnote{See e.g. Theorem 4.5.5. in \cite{Vershynin}.}.}} \textit{Assume $\mathcal{M}_0\subset \mathcal{M}$ and $ \mathcal{M}_1 \subset \mathcal{M}$} while $\mathcal{M}_0$ and $ \mathcal{M}_1$ are not necessarily equal nor subsets of each other. There exists the following in relation to the eigen-vectors $\mathscr{v}$ of $\mathcal{M}_0$ and $ \mathcal{M}_1$
\begin{equation}
\begin{split}
\;  sin \measuredangle \big(    \mathscr{v}_i(\mathcal{M}_0), \mathscr{v}_i(\mathcal{M}_1)  \big) \le \frac{2}{\gamma_x}||\mathcal{M}_0-\mathcal{M}_1||, \\
\gamma_x >0 \; \stackrel{def}{=}  \mathop{{\rm \mathbb{M}in}}\limits_{j \neq i} {\rm \; }|\zeta_i(\mathcal{M}_1)-\zeta_j(\mathcal{M}_0)|,\;\;\;\;\;\;\;\;\;\;
\end{split}
\end{equation}
while $\gamma_x >0$ is defined as the least separation distance of the largest eigen-value(s) from the rest of the spectrum.

The proof is now completed.$\; \; \; \blacksquare$

\section{Proof of Proposition 5}
\label{sec:H}
If the attack is a denial-of-service one and if it is perodic, as fully discussed e.g. in \cite{end1}, there consequently exist two totally i.i.d and separate scenraios in termf of two separately unstable and stable sub-systems between which there is a switching case. Now regarding the facts that: 
\begin{itemize}
\item (\textit{i}) the switching case theoretically entails a semi-Markov model\footnote{See e.g. \cite{end2} to understand the randomness of the time transitions and the necessity of semi-Markov modeling.}; and 
\item (\textit{ii}) the frequency of the occurrence in relation to our Markov model may be unavailable, that is, our knowledge of the information is somehow incomplete, the probability-theory is totally inappropriate\footnote{See e.g. \cite{end3, end4, end5, end6} to understand the differences between possibility-theory and probability-theory.} here, we consequently need to extend the probability transitions to the possibilities ones $-$ where the transitions stand for the strength and casuality $-$ and from a possibility-theoretic point of view;
\end{itemize}
one can extend the Markov model discussed in the previous part as below.

Provisionally speaking, re-call $\mathcal{P} \big(\zeta^{(\cdot)}_{max},\zeta^{(\cdot)}_{min} \big)$ from Appendix \ref{sec:G}. Now, if we are supposed to extend the Markov process $\mathcal{F} \big(\zeta_{set} \big), \; (\cdot)_{set} \in \{min, max\}$ to a semi-Markov one, while the timing jumps are randomly distributed as well, that is, $\mathcal{F} \big(\zeta_{set} \big)$ is only valid as follows
\begin{equation}
\begin{split}
\begin{cases}
 \mathcal{F} \big(\zeta_{set} (t) \big), \forall t_s \le t < \forall t_{s+1},\\
\mathscr{p}_{ij}\stackrel{def}{=}\mathbb{P}\mathscr{r} \big(\zeta^{(s+1)}_{set}=j, t_{s+1}-t_s \le t | \zeta^{(s)}_{set}=i \big)=\\ \;\;\;\;\;\;\;\; \; \;  \mathbb{P}\mathscr{r} \big(\zeta^{(s+1)}_{set}=j | \zeta^{(s)}_{set}=i \big),
\end{cases}
\end{split}
\end{equation}
while $(\cdot)_s$ stands for the $s-$th state.

\textsc{\textbf{Remark 8 $-$ An overview over the possibility theory \cite{end3}-\cite{Mei}.}} \textit{The following main rules hold in the possibility-theory: (i) the normality axiom indicates that $\upsilon(s)=1, \forall s \in \mathcal{S}$ holds; (ii) the non-negativity axiom indicates that $\upsilon(\varnothing)=0$; (iii) degree of possibility is derived by $\Upsilon(\mathcal{S})=\mathop{{\rm \mathbb{S}up}}\limits_{s\in \mathcal{S}} \upsilon(s)$; (iv) degree of possibility is derived by $\Upsilon^{(nec)}(\mathcal{S})=1-\mathop{{\rm \mathbb{I}nf}}\limits_{s \notin \mathcal{S}} \upsilon(s)$; (v) the maxitivity axiom\footnote{For example, if a person is a $50$-year-old one, if with the confidence of $1$ we say he/she is an ''aged'' person, $0.5$ a ''middle-aged'' person, and $0$ a ''young'' one, with the confidence of $1$ we can undoubtedly declare that he/she is ''adult''.} says that $\Upsilon(\mathcal{S}_1 \cup \mathcal{S}_2)=max \big \lbrace   \Upsilon(\mathcal{S}_1) , \Upsilon(\mathcal{S}_2) \big \rbrace, \forall \mathcal{S}_1, \mathcal{S}_2 \subseteq \mathcal{S}$; (vi)} the minitivity axiom says that $\Upsilon(\mathcal{S}_1 \cap \mathcal{S}_2)=min \big \lbrace   \Upsilon(\mathcal{S}_1) , \Upsilon(\mathcal{S}_2) \big \rbrace, \forall \mathcal{S}_1, \mathcal{S}_2 \subseteq \mathcal{S}$. Furthermore, the following conditional properties are information-theoretically satisfied \cite{Mei}
\begin{equation}
\begin{split}
\upsilon(s^{(i)}_1,s^{(j)}_2)=\upsilon(s^{(j)}_2|s^{(i)}_1)\upsilon(s^{(i)}_1),\;\;\;\;\;\;\;\;\;\;\;\;\;\;\;\;\;\;\;\;\;\;\\
\upsilon(s^{(i)}_1)=\frac{1}{\varpi_1(s^{(i)}_1)}\mathop{{\rm \mathbb{M}ax}}\limits_{s^{(j)}_2}\upsilon(s^{(i)}_1,s^{(j)}_2),\;\;\;\;\;\;\;\;\;\;\\
\upsilon(s^{(j)}_2)=\frac{1}{\varpi_2(s^{(j)}_2)}\mathop{{\rm \mathbb{M}ax}}\limits_{s^{(i)}_1}\upsilon(s^{(i)}_1,s^{(j)}_2),\;\;\;\;\;\;\;\;\;\\
\varpi_1(s^{(i)}_1)=\mathop{{\rm \mathbb{M}ax}}\limits_{s^{(j)}_2}\upsilon(s^{(j)}_2|s^{(i)}_1)\le 1,\;\;\;\;\;\;\;\;\;\;\;\;\;\;\;\;\;\;\\
\varpi_2(s^{(j)}_2)=\mathop{{\rm \mathbb{M}ax}}\limits_{s^{(i)}_1}\upsilon(s^{(i)}_1|s^{(j)}_2)\le 1,\;\;\;\;\;\;\;\;\;\;\;\;\;\;\;\;\;\;\\
\upsilon(s^{(i)}_1|s^{(j)}_2)=\frac{\upsilon(s^{(i)}_1)\upsilon(s^{(j)}_2|s^{(i)}_1)}{\upsilon(s^{(j)}_2)}\;\;\;\;\;\;\;\;\;\;\;\;\;\;\;\;\;\;\;\;\;\\
=\frac{\varpi_2(s^{(j)}_2)\upsilon(s^{(i)}_1)\upsilon(s^{(j)}_2|s^{(i)}_1)}{\mathop{{\rm \mathbb{M}ax}}\limits_{s^{(k)}_1} \big \lbrace    \upsilon(s^{(k)}_1)\upsilon(s^{(j)}_2|s^{(k)}_1)   \big   \rbrace},\;\;\;\;\;\;\;\\
\end{split}
\end{equation}
as well as \cite{Mei}
\begin{equation}
\begin{split}
\upsilon(s^{(k)}_3|s^{(i)}_1)=\frac{1}{\varpi_3(s^{(i)}_1)}\mathop{{\rm \mathbb{M}ax}}\limits_{s^{(j)}_2} \upsilon(s^{(k)}_3|s^{(j)}_2)\upsilon(s^{(j)}_2|s^{(i)}_1),\\
\varpi_3(s^{(i)}_1)=\mathop{{\rm \mathbb{M}ax}}\limits_{s^{(j)}_2}\upsilon(s^{(j)}_2|s^{(i)}_1)\le 1,\;\;\;\;\;\;\;\;\;\;\;\;\;\;\;\;\;\;\;\;\;\;\;\;\;
\end{split}
\end{equation}
and \cite{Mei}
\begin{equation}
\begin{split}
\mathop{{\rm \mathbb{M}ax}}\limits_{s^{(j)}_2}   \upsilon(s^{(k)}_3, s^{(j)}_2|s^{(i)}_1)=\;\;\;\;\;\;\;\;\;\;\;\;\;\;\;\;\\\mathop{{\rm \mathbb{M}ax}}\limits_{s^{(j)}_2}  \upsilon(s^{(k)}_3|s^{(j)}_2, s^{(i)}_1) \upsilon(s^{(j)}_2|s^{(i)}_1)  =\\ \mathop{{\rm \mathbb{M}ax}}\limits_{s^{(j)}_2}\upsilon(s^{(k)}_3|s^{(j)}_2) \upsilon(s^{(j)}_2|s^{(i)}_1),
\end{split}
\end{equation}
which is also equal to \cite{Mei}
\begin{equation}
\begin{split}
\mathop{{\rm \mathbb{M}ax}}\limits_{s^{(j)}_2}  \upsilon(s^{(j)}_2|s^{(k)}_3, s^{(i)}_1) \upsilon(s^{(k)}_3|s^{(i)}_1)  =\\ \mathop{{\rm \mathbb{M}ax}}\limits_{s^{(j)}_2} \upsilon(s^{(j)}_2|s^{(i)}_1) \upsilon(s^{(k)}_3|s^{(i)}_1) =\\
\varpi_3(s^{(i)}_1)\upsilon(s^{(k)}_3|s^{(i)}_1).\;\;\;\;\;\;\;\;\;\;\;\;\;
\end{split}
\end{equation}

In addition, the following equations are also added \cite{Benferhat1, Benferhat2}
\begin{equation}
\begin{split}
 \upsilon(s_1|s_2 \cdots, s_N) = 
\begin{cases}
\frac{\upsilon(s_1, \cdots, s_N)}{\upsilon(s_2, \cdots, s_N)}, \;\upsilon(s_2, \cdots, s_N) \neq 0,\\
1, \;\;\;\;\;\;\;\;\;\;\; \;\;\;\;\upsilon(s_2, \cdots, s_N) = 0,
\end{cases}
\end{split}
\end{equation}
as well as
\begin{equation}
\begin{split}
 \upsilon(s_1, \cdots, s_N) = \prod_{i}\upsilon \big(s_i|\mathscr{Par} (s_i)\big),
\end{split}
\end{equation}
while the parent elements\footnote{Casual prior samples.} $\mathscr{Par} (s_i)$ are the ones defined by the Cartesian poruct of the main set's domain, aacording to the following definition for the possibilistic graphs.

\textsc{\textbf{Definition 5: Possibilistic graph \cite{Benferhat1, Benferhat2}.}} \textit{A possibilistic casual network is defined in terms of the graph $\mathscr{G}_{\mathscr{poss}}\stackrel{def}{=}\bigg \lbrace \Big(   s, \mathscr{Par} (s), \alpha_{\mathscr{poss}}   \Big): \upsilon \big( s| \mathscr{Par} (s)\big)=\alpha_{\mathscr{poss}}  \neq 1     \bigg \rbrace$.}

The proof is now completed.$\; \; \; \blacksquare$

\markboth{IEEE, VOL. XX, NO. XX, X 2021}%
{Shell \MakeLowercase{\textit{et al.}}: Bare Demo of IEEEtran.cls for Computer Society Journals}
\end{document}